\documentclass{article} 
\usepackage[dvips]{graphicx}
\def\be{\begin{eqnarray}}
\def\ee{\end{eqnarray}}

\def\abstract#1{\vskip 7mm 
\begin{center}{\large Abstract}\par \bigskip
\begin{minipage}[c]{12cm}
\small #1
\end{minipage}
\end{center}
}
\def\title#1{\begin{center}{\Large\bf #1}\end{center}}
\def\author#1{\vskip 5mm \begin{center}{#1}\end{center}}
\def\address#1{\begin{center}{\it #1}\end{center}}

\newcommand{\bfr}{\begin{flushright}}
\newcommand{\efr}{\end{flushright}}
\begin{document} 
\vspace*{-0cm}
\title{Lie Algebra Quantization by the Star Product}
\author{Takao KOIKAWA\footnote{E-mail: koikawa@otsuma.ac.jp}
}
\vspace{1cm}
\address{
${}$ School of Social Information Studies,\\
        Otsuma Women's University\\
        Tama 206-0035,~Japan\\
}
\vspace{5.5cm}
\abstract{
We apply the star product quantization to the Lie algebra. The quantization in terms of the star  product is well known and the commutation relation in this case is called the $\theta$-deformation where the constant $\theta$ appears as a parameter. In the application to the Lie algebra, we need to change the parameter $\theta$ to $x$-dependent $\theta(x)$. There is no essential difference between the quantization in the quantum mechanics and deriving quantum numbers in the Lie algebra from the viewpoint of the star product. We propose to unify them in higher dimensions, which may be analogous to the Kaluza-Klein theory in the classical theory.}

\newpage
\setcounter{page}{2}
\noindent
%
%
\section{introduction}
The commutation relations(CRs hereafter) of non-commutative coordinates $\hat x^i$ can be classified into several types. Two of the important types of the CRs of operators $\hat x^i$ are given by\cite{Mey}
\be
\left[{\hat x}^i,{\hat x}^j\right]&=&i\Theta^{ij},\label{typ1}\\
\left[{\hat x}^i,{\hat x}^j\right]&=&i\theta f_k^{~ij}\hat x^k,\label{typ2}
\ee
where $\Theta^{ij}$, $f_k^{~ij}$ and $\theta$ are constants. The first CR is the $\theta$-deformation type, and the second CR is the Lie algebra type.

The quantization using the star product is mainly applied to the first type. The application of the star product to the second type is less known. It might be possible to unify these types as
\be
\left[{\hat x}^i,{\hat x}^j\right]&=&i\Theta^{ij}(\hat x),\label{typ23}
\ee
where $\Theta^{ij}(\hat x)$ are functions of ${\hat x}^i$s in general. A special case $\Theta^{ij}(\hat x)=\Theta^{ij}$ falls into the first type. So far, $\theta$-deformation known as the Moyal quantization has been studied extensively. On the other hand, the space-time operator dependent  $\Theta^{ij}(\hat x)$ case has been less studied from the viewpoint of star product framework.

One of the purposes of the present paper is to show the star product realization of the Lie algebra type of operator CRs. When $f_k^{~ij}$ in the CR (\ref{typ2}) is the structure constant of $su(2)$, the star product representation of the CR also yields the algebra in the same way as the operator representation. The quantization of the Casimir operator is carried out by using functions and star product only. 

When we derive a star product for the Lie algebra type CR, we encounter a difficulty which does not exist in the $\theta$-deformation. In order to clarify the difficulty we start with reviewing the definition of the star product for the  $\theta$-deformation case. We define the star product for functions of $x^1$ and $x^2$, $f(x)=f(x^1, x^2)$ and $g(x)=g(x^1, x^2)$, by
\be
&&f(x)\star g(x)\cr
&=&\exp\left[i\frac{\theta}{2}\left(\frac{\partial}{\partial x^1}\frac{\partial}{\partial {x^2}'}-\frac{\partial}{\partial x^2}\frac{\partial}{\partial{x^1}'}\right)\right]f(x) g(x')|_{{x^1}'=x^1,~{x^2}'={x^2}}, \label{thetadfm}
\ee
where $\theta$ is a constant. As far as $\theta$ is a constant, the differential operators in the exponent never hit $\theta$ when the exponential function is expanded. However, when $\theta$ is a function of $x^1$ and $x^2$, there occurs a question whether the $x$-dependent $\theta$ is differentiated by the differential operators in the exponent of the exponential function. Though there can be several definitions, we adopt the ansatz that $\theta(x)$ is not differentiated. In other words, we generalize the above definition to the following definition for $x$-dependent $\theta(x)$ case. We denote the star product by use of the same notation as before
\be
&&f(x)\star g(x)\cr
&=&\exp\left[i\frac{\theta (x'')}{2}
\left(\frac{\partial}{\partial x^1}\frac{\partial}{\partial {x^2}'}-\frac{\partial}{\partial x^2}\frac{\partial}{\partial{x^1}'}\right)
\right]f(x) g(x')|_{{x^1}''={x^1}'=x^1,~{x^2}''={x^2}'={x^2}}\cr    
&=&\sum_{n=0}^{\infty} \frac{1}{n!}\left(i\frac{\theta (x'')}{2}\right)^n\cr
&& \quad \times \left(\frac{\partial}{\partial x^1}\frac{\partial}{\partial {x^2}'}-\frac{\partial}{\partial x^2}\frac{\partial}{\partial{x^1}'}\right)^nf(x) g(x')|_{{x^1}''={x^1}'=x^1,~{x^2}''={x^2}'={x^2}}.\label{xdeptheta}
\ee
The last expression is to be read that the Poisson bracket differential operator(see Eq. (\ref{Poisson})) acts upon $f$ and $g$ only, and we keep this ansatz whenever we define other type of star products throughout this paper.

This paper is organized as follows. In the following section, we discuss the $x$-dependent $\theta$-deformation. We show how such cases appear in curved space-time. In section 3, we review the $\theta$-deformation quantization of one-dimensional harmonic oscillator which shows one of the most general procedure of Moyal quantization. In section 4, we quantize the Lie algebra $su(2)$ and $su(3)$ using the star product in the same way as in the Moyal quantization, after we assign proper components to each element of a matrix $(\Theta^{\mu\nu}(x))$  for the Lie algebra. The quantization is implemented completely in an algebraic way.  In section 5, we propose a higher dimensional model where the quantization of a Hamiltonian and the quantization of the Lie algebra in terms of the star product are unified in higher dimensions. The last section is devoted to summary and discussion.
 
%
%
\section{Generalization of the Star Product}
The procedure of defining the star product for constant $\theta$ is a little bit cumbersome as shown in Eq.(\ref{thetadfm}). First we need to distinguish the variables of a function $f(x)$ and $g(x)$ and then take derivatives by allocating the variables to derivative operators. After the derivatives are taken, the variables are set to be equal.  

When we change the constant $\theta$ to a function of $x^1$ and $x^2$, $\theta(x)$, manipulation becomes more complicated. As for a question whether $\theta(x)$ is differentiated by the differential operators in the exponent, we adopt the ansatz that the differential operators operate only on the functions $f(x^1, x^2)$ and $g(x^1, x^2)$ and not on $\theta(x)$ as in Eq.(\ref{xdeptheta}). The merit of this ansatz is that the Poisson bracket operator defined by
\be
\frac{\partial}{\partial x^1}\frac{\partial}{\partial {x^2}}-\frac{\partial}{\partial x^2}\frac{\partial}{\partial{x^1}},  \label{Poisson}
\ee
on the functions always keeps its form because part of the operators never operates on $\theta(x)$,  by assumption. We keep this ansatz as a prescription for the generalization of the star product in the discussion below. We also comment on the reason why we need to study $x$-dependent $\theta(x)$.

In order to generalize the star product, we replace differential operators by vector fields. Then $x$-dependent coefficient in front of the differential operators appear, and we follow the above ansatz in this case. The generalized star product is explicitly written as
\be
f(x)\star g(x)=\mu \left \{ \exp\left[i\frac{\Theta^{a b}}{2}X_{a}\bigotimes X_{b}\right]f(x)\bigotimes g(x)\right \}, \label{genstar}
\ee
where the independent vector field $X_a,~(a=1,2,\cdots,n)$ is given by
\be
X_a=e_a^{\mu}(x)\partial_{\mu},
\ee
and is  assumed to satisfy
\be
[X_a,~X_b]=0.
\ee
In (\ref{genstar}), $\Theta^{a b}~(a,~b=1,2,\cdots,n)$ is an antisymmetric constant in indices $a$ and $b$, and $f(x)$ and $g(x)$ are functions of $n$ variables $x^i,~(i=1,2,\cdots,n)$. The expansion of the generalized star product is written as
\be
&&f(x)\star g(x)\cr
&=&f(x)g(x)+\frac{i}{2}\Theta^{a b}X_af(x)X_bg(x) \cr
&&+\frac{1}{2!}\left(\frac{i}{2}\right)^2 \Theta^{a_1 b_1} \Theta^{a_2 b_2} (X_{a_1}X_{a_2}f(x))(X_{b_1}X_{b_2}g(x))\cr
&&+\frac{1}{3!}\left(\frac{i}{2}\right)^3 \Theta^{a_1 b_1} \Theta^{a_2 b_2} \Theta^{a_3 b_3}(X_{a_1}X_{a_2}X_{a_3}f(x))(X_{b_1}X_{b_2}X_{b_3}g(x))+\cdots \cr
&=&f(x)g(x)+\frac{i}{2}\Theta^{\mu \nu}\partial_{\mu}f(x)\partial_{\nu}g(x)\cr
&&+\frac{1}{2!}\left(\frac{i}{2}\right)^2 \Theta^{\mu_1 \nu_1} \Theta^{\mu_2 \nu_2} (\partial_{\mu_1}\partial_{\mu_2}f(x))(\partial_{\nu_1}\partial_{\nu_2}g(x))\cr
&&+\frac{1}{3!}\left(\frac{i}{2}\right)^3 \Theta^{\mu_1 \nu_1} \Theta^{\mu_2 \nu_2} \Theta^{\mu_3 \nu_3}(\partial_{\mu_1}\partial_{\mu_2}\partial_{\mu_3}f(x))(\partial_{\nu_1}\partial_{\nu_2}\partial_{\nu_3}g(x))+\cdots,
\ee
where the coefficients $\Theta^{\mu \nu}$ appearing in front of the differential operators are functions of $x$ given by
\be
\Theta^{\mu \nu}=\Theta^{\mu \nu}(x)=\Theta^{a b}e_a^{\mu}(x)e_b^{\nu}(x).
\ee
Here we have followed the prescription that the differential operators operate only on the functions $f(x)$ and $g(x)$, and not on $\Theta^{\mu \nu}(x)$. We thus define the generalize star product by
\be
&&f(x)\star g(x)\cr
&=&\sum_{n=0}^{\infty}\frac{1}{n!}\left(i \frac{\Theta^{\mu\nu}(x'')}{2}\right)^n\left(\frac{\partial}{\partial x^{\mu}}\frac{\partial}{\partial {x^{\nu}}'}-\frac{\partial}{\partial x^{\nu}}\frac{\partial}{\partial{x^{\nu}}'}\right)^nf(x) g(x')|_{x''=x'=x}. \label{genstarprd}
\ee
We note that this formula of the generalized star product coincides with the ordinary definition of the star product (\ref{xdeptheta}) when $\Theta^{\mu \nu}=\theta \epsilon^{\mu \nu}$. We stress that we obtain this formula similar to one with a constant coefficient, as the consequence of the prescription, otherwise there can appear terms which include the derivatives of vielbein $e_a^{\mu}(x)$s.

We denote the star product CR by
\be
[f(x) ,g(x)]_\star=f(x)\star g(x)-g(x)\star f(x).
\ee
We apply the formula to the case $f(x)=x^{\mu}$ and $g(x)=x^{\nu}$ to obtain 
\be
[x^{\mu},x^{\nu}]_\star= i\Theta^{\mu \nu}(x), \label{qoe}
\ee
where $\mu,~\nu=1,2,\cdots,n$. In this paper, we do not take up the form of vielbein too much. We rather take this equation as a starting point.  By assuming possible forms of $\Theta^{\mu \nu}(x)$, we study what we can obtain from this equation as the result of possible assumptions.

%
%
\section{One-Dimensional Harmonic Oscillator}
In this section, we quantize the one-dimensional harmonic oscillator by using the star product. The energy spectrum is quantized using the stargenvalue equation\cite{HK}. The one-dimensional harmonic oscillator falls into an application of the generalized star product by inserting $\Theta^{\mu \nu}(x)=\theta \epsilon^{\mu \nu},~(\mu,~\nu=1,2)$ with constant $\theta$, which is the simplest case of the generalized star product cases, or the ordinary star product used for the $\theta$-deformation. Inserting this into Eq.(\ref{qoe}), we obtain
\be
[x^1,x^2]_\star=i\theta.
\ee
 The star product which we use in this section is obtained from (\ref{genstarprd}) by setting $\Theta^{\mu \nu}(x)=\theta \epsilon^{\mu \nu}$. The dimension is two which is the smallest dimension in the phase space of coordinate and momentum. As far as we apply the star product to the $\theta$-deformation to such phase space, the dimension is always even. This example shows contrasting features in comparative with the Lie-algebra types discussed in the following section, which allow for the odd number of dimensions.

We introduce new variables  $a$ and $\bar a$ by
\be
a=\frac{x^1+ix^2}{\sqrt 2},\\
\bar a=\frac{x^1-ix^2}{\sqrt 2}.
\ee
Then the differentiations with respect to $x^1$ and $x^2$ are replaced by those with respect to $a$ and $\bar a$ as
\be
\frac{\partial}{\partial x^1}&=&\frac{1}{\sqrt 2}\left(\frac{\partial}{\partial a}+\frac{\partial}{\partial \bar a}\right),\\
\frac{\partial}{\partial x^2}&=&i\frac{1}{\sqrt 2}\left(\frac{\partial}{\partial a}-\frac{\partial}{\partial \bar a}\right).
\ee
Inserting these results to the star product, we obtain
\be
f\star g=\exp\left[\frac{\theta}{2}\left(\frac{\partial}{\partial a}\frac{\partial}{\partial \bar a'}-\frac{\partial}{\partial a}\frac{\partial}{\partial\bar a'}\right)\right]f(a,\bar a) g(a',\bar a')|_{a'=a,~\bar a'=\bar a'}.
\ee
Then, the CR is written in terms of $a$ and $\bar a$ as
\be                                
[a,\bar a]_{\star}=\theta,
\ee
suggesting $a$ and $\bar a$ play roles of an annihilation operator and a creation operator in the operator formalism, respectively.

We next discuss the stargen equation of the one-dimensional harmonic oscillator  Hamiltonian given by
\be
H=\frac{1}{2}((x^1)^2+(x^2)^2)=\frac{1}{2}(\bar a\star a+a\star \bar a).
\ee
We shall study the stargenstates belonging to the stargenvalues of the Hamiltonian. The stargenstate $\psi$ and stargevalue $E$ correspond to the eigenstate and eigenvalues in the operator formalism, respectively. The stargen-value equation is given by
\be
H\star \psi=E\psi. \label{stargeneq}
\ee
In order to solve this equation, there can be an analytic method and an algebraic method. Here we use the algebraic method. We first rearrange the order of $a$ and $\bar a$ in the Hamiltonian as
\be
H=\bar a \star a+\frac{\theta}{2}=a \star \bar a-\frac{\theta}{2}=\frac{1}{2}(a \star \bar a+\bar a \star a).
\ee
We then define the vacuum state $\psi_0$ by
\be
a\star \psi_0=0.
\ee
The stargen-value equation is solved by
\be
H\star \psi_n=E_n\psi_n,~(n=0,1,2\cdots) \label{sol1}
\ee   
where the stargenstate $\psi_n$ and the stargen-value $E_n$ are given by
\be
\psi_n=(\bar a\star)^n\psi_0,~E_n=(n+\frac{1}{2})\theta.   \label{sol2}
\ee
This is proved by the inductive method. The vacuum state satisfies 
\be
H\star \psi_0=(\bar a \star a+\frac{\theta}{2})\star \psi_0=\frac{\theta}{2}\star \psi_0,
\ee
showing that Eq.(\ref{sol2}) holds for $n=0$. We assume that Eq.(\ref{sol1}) holds for $n=k$. When $n=k+1$, the stargen-value equation reads
\be
H\star \psi_{k+1}=H\star \bar a \star \psi_{k}=(\bar a\star H+\theta \bar a)\star \psi_k=(k+1+\frac{1}{2})\theta \psi_{k+1},
\ee
where we have used
\be
[H,\bar a]_\star=\theta \bar a.
\ee
Therefore, Eq.(\ref{sol1}) holds also for $n=k+1$. We have thus shown that Eq.(\ref{stargeneq}) is solved by Eqs.(\ref{sol1}) with Eq.(\ref{sol2}) for non-negative integer number $n$.

This is a well-known result in the operator formalism or the Schr\"{o}dinger equation, and we have shown that the same result can be obtained by solving the stargen-value equation.  We remark here that the energy spectrum is semi-infinite.

%
%
\section{Lie algebra}
In this section, examples of $x$-dependent $\Theta^{\mu \nu}(x)$ in (\ref{qoe}) are shown. In the first subsection, we show $su(2)$ as an example. In the second subsection, $su(3)$ is exhibited as another example.
\subsection{su(2)}
We present a three dimensional space model in which the coordinates $x^{\mu}$ satisfy Eq.(\ref{qoe}) with a condition $n=3$:
\be
[x^{\mu},x^{\nu}]_{\star}=i\Theta^{\mu \nu}(x), ~(\mu,~\nu=1,2,3) \nonumber
\ee
where $\Theta^{\mu \nu}(x)$ is an entity of a $3\times 3$ matrix given by
\be
(\Theta^{\mu \nu}(x))=2\theta
\pmatrix{
0 	& x^3 	& -x^2\cr
-x^3 	& 0	& x^1\cr
x^2	&-x^1	& 0
}.
\ee
Then, Eq.(\ref{qoe}) with this assignment yields the $su(2)$ CR in terms of the star product
\be
[x^{\mu},x^{\nu}]_\star=2i\theta \epsilon_{\mu \nu \lambda} x^{\lambda},\label{so3}
\ee
where $\epsilon_{\mu \nu \lambda}$ are constants antisymmetric in all indices with $\epsilon_{123}=1$. 

We next discuss the quantization of this algebra. Introducing $j_{\pm}$ and $j_3$ by
\be
j_{\pm}&=&\frac{x^1\pm ix^2}{2},\\
j_3&=&\frac{x^3}{2},
\ee
the CRs are rewritten as 
\be
[j_3,j_{\pm}]_{\star}&=&\pm \theta j_{\pm},\label{creation}\\
\left[j_{+},j_{-}\right]_{\star}&=&2 \theta j_3.
\ee
We define $j^2$ known as a Casimir operator in the operator formalism by
\be
j^2=\sum_{i=1}^3j_i^2,
\ee
then the CRs with $j$s become
\be
\left[j_3,j^2\right]_{\star}&=&0,\\
\left[j_{\pm},j^2\right]_{\star}&=&0,
\ee 
which show that $j^2$ is commutable with all $j$s. 

In order to obtain stargen-values and stargen functions, we need to solve the stargen-value equations of $j_3$ and $j^2$. Since they are commutable, the stargen functions are characterized by two indices which are independent to each other. We denote them by $l$ and $m$. In the quantization of a harmonic oscillator in the previous section, we quantized the energy spectrum of the Hamiltonian function. In the present case, we quantize the $su(2)$ algebra in which the stargen function is that of $j_3$ and $j^2$ at the same time. This may be similar to the situation where a constraint on the eigenstate of a Hamiltonian is imposed.

We study the stargen-value equation of $j_3$. We define the state belonging to the stargen-value $m \theta$ by
\be
j_3\star f_{lm}=m \theta f_{lm},
\ee
where $m$ is assumed to be an integer or half-integer. 
As for the range of $m$, a condition which bounds the range will be imposed later. We also assume
\be
j^2\star f_{lm}=\gamma \theta^2 f_{lm},
\ee
where $\gamma$ is assumed to be expressed in terms of $l$ only and its explicit form will also be computed later.

The states $f_{lm\pm k}$ with the second quantum number $m\pm k$ can be obtained by
operating $j_{\pm}$ upon  $f_{lm}$ state $k$ times as
\be
f_{lm\pm k}\propto (j_{\pm}\star)^kf_{lm}.\label{excited}
\ee
These can be derived in the following way. By using (\ref{creation}), we obtain
\be
j_3\star (j_{\pm}\star f_{lm})=(j_{\pm}\star j_3{\pm}\theta j_{\pm})\star f_{lm}=(m{\pm}1)\theta (j_{\pm}\star f_{lm}).
\ee
Then we find that
\be
j_{\pm}\star f_{lm}\propto f_{lm{\pm}1}.
\ee
By repeating this procedure,  Eq.(\ref{excited}) can be proved.

As for the range of $m$, we assume that $m$ ranges from $-l$ to $l$ where we assume that $l$ is integer or half-integer. In order this to be realized, we impose a condition that the value $m$ does not exceed the maximal value $l$:
\be
j_+\star f_{ll}=0.
\ee
By taking a complex conjugate of this equation, we also have
\be
f_{l-l} \star j_{-}=0.
\ee
These conditions guarantee the finite range of $m$;$-l\le m \le l$.

Last of all, we determine the explicit form of $\gamma$, which is expressed in terms of $l$. Noting that
\be
j^2=j_-\star j_++\theta j_3+j_3^2,
\ee
we obtain, by operating $j^2$ on $f_{ll}$, 
\be
j^2\star f_{ll}=(j_-\star j_++\theta j_3+j_3\star j_3)\star f_{ll}=(l^2+l)\theta^2f_{ll}.
\ee
Then, by operating $j_-$ on $f_{ll}$ $(l-m)$ times, we obtain
\be
j^2\star f_{lm}=(l^2+l)\theta^2 f_{lm}.
\ee
We have thus derived that $\gamma=l(l+1)$.

We have assumed that $l$ takes either the integer value or half-integer value. When $l$ takes integer values, the algebra represents $so(3)$. Setting $\theta=\hbar$, we obtain the algebra of the angular momentum. When $l$ takes half-integer values, the algebra represents $su(2)$, which can be interpreted as a spin representation or the internal symmetry, isospin.  For example, when $l$ is set to be $l=1/2$, the value of $m$ takes either $m=1/2$ or $m=-1/2$, which correspond to an upstate and a downstate of isospin, respectively.
\subsection{su(3)}
In this subsection, we discuss $su(3)$ algebra in terms of the star product. We start with defining the star product. The rank of the root space of $su(3)$ is two in contrast to one in $su(2)$ case. This makes it complicated to impose constraints in order to confine the states into the multiplet representation of $su(3)$. 

We study a eight dimensional space model in which the coordinates $x^i,~(i=1,2,\cdots,8)$ satisfy Eq.(\ref{qoe}) with a condition $n=8$:
\be
 [x^{\mu},x^{\nu}]_{\star}=2if_{\mu \nu \lambda}x^{\lambda}=i\Theta^{\mu \nu}(x), \nonumber
\ee
where $f_{\mu \nu \lambda}$ is the $su(3)$ structure constant and $\Theta^{\mu \nu}(x)$ is given by
\be
&&(\Theta^{\mu \nu}(x)) \cr 
&=& \theta \pmatrix{
0           & 2x^3 & -2x^2 & x^7 & -x^6 & x^5 & -x^4 & 0 \cr
-2x^3 & 0 & 2x^1 & x^6 & x^7 & -x^4 & -x^5 & 0 \cr
2x^2  & -2x^1 & 0 & x^5 & -x^4 & -x^7 & x^6 & 0 \cr
-x^7 & -x^6 & -x^5 & 0 & x^3 & x^2 & x^1 & -\sqrt 3x^5 \cr
x^6 & -x^7 & x^4 & -x^3 & 0 &  -x^1 & x^2 &  \sqrt 3x^4 \cr
-x^5 & x^4 & x^7 & -x^2 & x^1 & 0 & -x^3 & -\sqrt 3x^7 \cr
x^4 & x^5 & -x^6 & -x^1 & -x^2 & x^3 & 0 & \sqrt 3x^6 \cr
0&0 & 0 & \sqrt 3x^5 & -\sqrt 3x^4 &  \sqrt 3x^7 & -\sqrt 3x^6 & 0  
}.\nonumber 
\ee
For example, we can read the following examples from this matrix
\be
[x^1,x^2]_{\star}=2i\theta x^3,~[x^3,x^1]_{\star}=2i\theta x^2,~[x^2,x^3]_{\star}=2i\theta x^1,
\ee
which show that the CRs constitute one of the $su(2)$ subalgebras of $su(3)$.

We introduce new variables in order to adapt to the ordinary expressions. We define $H_i,~(i=1,2)$ by
\be
H_1&=&\frac{x^3}{2},\\
H_2&=&\frac{1}{\sqrt 3}x^8,
\ee
which are commutable with each other
\be
[H_1,~H_2]_{\star}=0.
\ee
We further define linear combinations of $x^i$ by
\be
E_{\pm 1}= (x^1 \pm ix^2)/2, E_{\pm 2}= (x^6 \pm ix^7)/2,~E_{\pm 3}=(x^4 \pm ix^5)/2.
\ee
Some of the CRs are given by
\be
\left[E_{1},E_{-1}\right]_{\star}&=&2H_{1},\\
\left[E_{2},E_{-2}\right]_{\star}&=&\frac{3}{2}H_{2}-H_{1},\label{CR1}\\
\left[E_{3},E_{-3}\right]_{\star}&=&\frac{3}{2}H_{2}+H_{1}.
\ee
We introduce two component vector $\vec H$ by
\be
\vec H=\pmatrix{H_1 \cr H_2}.
\ee
Then we obtain
\be
\left[\vec H, E_{\pm i}\right]_{\star}=\pm \vec \alpha_{i} E_{\pm i},~(i=1,2,3)
\ee
where two component vectors $\vec \alpha_i$s called root vectors are given by
\be
\vec \alpha_1=\pmatrix{1 \cr 0},~\vec \alpha_2=\pmatrix{-\frac{1}{2} \cr 1},~\vec \alpha_3=\pmatrix{\frac{1}{2} \cr 1}=\vec \alpha_1+\vec \alpha_2.
\ee

We next study the stargen-value equation of $\vec H$. We denote the stargen function of $\vec H$ by $f_{I_3 Y}(x)$, where $I_3$ and $Y$ are the stargen-values of $H_1$ and $H_2$, respectively. The equation reads
\be
\vec H \star f_{I_3 Y}(x)=f_{I_3 Y}(x) \star \vec H=\vec \eta f_{I_3 Y}(x),
\ee
where $\vec \eta=(I_3,Y)^t$. Though there are left and right operation of $\vec H$ in the stargen-value equation, the operation of $\vec H$ from left shall be omitted in the following discussion for simplicity, if not necessary.

\begin{figure}
\centerline{\includegraphics[width=5cm,height=5cm]{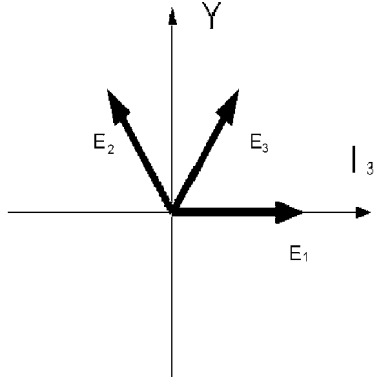}}
\caption{The translations by $E_1$, $E_2$ and $E_3$ expressed by $\vec \alpha_1$, $\vec \alpha_2$ and $\vec \alpha_3$. }
\label{fig:1}
\end{figure}

We study the translations of a point specified by the stargen-values $(I_3,Y)$ in $I_3Y$-plane by the operation of $E_{\pm  i}$s. This can be explicitly shown by
\be
\vec H \star (E_{\pm i}\star f_{I_3 Y}(x))=(\vec \eta \pm \vec \alpha_{\pm i})(E_{\pm i}\star f_{I_3 Y}(x)).
\ee
This shows that the state $E_{\pm i}\star f_{I_3 Y}(x)$ is the stargen state belonging to the stargen-value ${\vec \eta} \pm {\vec \alpha}_{\pm i}$:
\be
E_{\pm i}\star f_{\vec \eta}(x) \propto f_{\vec \eta \pm \vec \alpha_i}(x),
\ee
where $f_{\vec \eta}(x)=f_{I_3 Y}(x)$. From this result, we find that $E_{\pm i}$ translate a state at $\vec \eta$ to one at $\vec \eta \pm \vec  \alpha_i$. The translations by $E_{\pm i},~(i=1,2,3)$ are illustrated in Fig.\ref{fig:1}.

\begin{figure}
\centerline{\includegraphics[width=5cm,height=5cm]{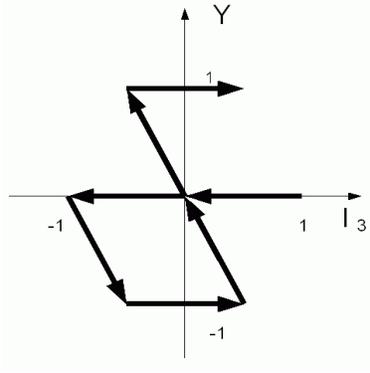}}
\caption{Traveling in the octet states driven by $E_{\pm i}$ starting from $f_{10}$ state.}
\label{fig:2}
\end{figure}

We next show that every state in a multiplet can be visited by these operations by $E_{\pm i}$. As an example, we can visit all the states in the octet starting from a state in the octet. Starting from a $f_{10}(x)$ state, other states can be visited by successive operations of $E_{\pm i}$s as
\be
f_{00}(x)&\propto&E_{-1}\star f_{10}(x),\\
f_{-10}(x)&\propto&E_{-1}\star f_{00}(x),\\
f_{-\frac{1}{2} -1}(x)&\propto&E_{-2}\star f_{-10}(x),\\
f_{\frac{1}{2} -1}(x)&\propto&E_{1}\star f_{-\frac{1}{2} -1}(x),\\
f_{00}(x)&\propto&E_{2}\star f_{\frac{1}{2} -1}(x),\\
f_{-\frac{1}{2} 1}(x)&\propto&E_{2}\star f_{00}(x),\\
f_{\frac{1}{2} 1}(x)&\propto&E_{1}\star f_{-\frac{1}{2} 1}(x),
\ee
which is shown in Fig.\ref{fig:2}.

When we observe the traveling in the octet in Fig.2, we note that the state designated by $(I_3,Y)=(0,0)$ is visited twice. As will be shown later, we have
\be
(E_{-1}\star )^2f_{00}(x)=0,\label{ker1}\\
(E_{2}\star )^2f_{00}(x)=0.\label{ker2}
\ee
Since $(E_{-1}\star )^2 \ne (E_{2}\star )^2$, this suggests that the state $f_{00}(x)$ is doubly degenerate, though it can not be distinguished by the values of $I_3$ and $Y$.

In the above traveling among the octet states, the path should not exceed the boundary of the hexagon. These constraints on the path should be imposed so that the traveling is limited inside an octet. After the introduction of the Casimir operator, we shall discuss these constraints. 

In $su(3)$ algebra, there are two Casimir operators. The stargen-values of the Casimir operators should be the same for any state in the octet, because they are commutable with all $x^i$s and so $E_i$s. We discuss one of the Casimir operators. The Casimir operator $C(su(3))$ is defined by
\be
C(su(3))=\sum_{i=1}^8 \left(\frac{x^i}{2}\right)^2.
\ee
Then, this can be rewritten in terms of $H_i$s and $E_{\pm i}$s as
\be
C(su(3))&=&H_1(H_1+1)+\frac{3}{4}H_2(H_2+2)+\sum_{i=1}^3E_{-i}\star E_i.
\ee
At each state on the vertices of hexagon representing the octet states, we should impose conditions that forbid the translations outward of the hexagon to the outer states and allow translations inward to the states within the hexagon. In order to realize this, we impose the following conditions at $(I_3,Y)=(1,0)$
\be
E_1*f_{10}(x)=0,~E_{-2}*f_{10}(x)=0,~E_3*f_{10}(x)=0.\label{con1}
\ee
At $(I_3,Y)=(\frac{1}{2},1)$, the conditions are given by
\be
E_1*f_{\frac{1}{2}1}(x)=0,~E_{2}*f_{\frac{1}{2}1}(x)=0,~E_3*f_{\frac{1}{2}1}(x)=0, \label{constr2}
\ee
 and at $(I_3,Y)=(-\frac{1}{2},1)$
\be
E_{-1}*f_{-\frac{1}{2}1}(x)=0,~E_{2}*f_{-\frac{1}{2}1}(x)=0,~E_3*f_{-\frac{1}{2}1}(x)=0.
\ee
As for other states at the antipodal vertices, similar conditions with $E_i$s with opposite signed $i$ are obtained by the complex conjugation of the above constraints. Noting that $f_{\frac{1}{2}1}(x)\propto E_2\star f_{00}$ and using the middle equation of (\ref{constr2}), we obtain
\be
(E_2\star)^2f_{00}=0,\nonumber
\ee
which is Eq.(\ref{ker2}). This shows that the state $f_{00}$ is a state of the triplet of $su(2)$. In the similar way, we can show Eq.(\ref{ker1}). Since there are two independent $su(2)$ subalgebras of $su(3)$, we have Eqs.(\ref{ker1}) and (\ref{ker2}) imposed on $f_{00}$ independently.

We can now evaluate the Casimir operator. As we have mentioned, evaluation at any state would be the same. We evaluate it at $(I_3,Y)=(1,0)$. We need to rewrite the Casimir operator so that the conditions (\ref{con1}) are made use of. By using the CR (\ref{CR1}), the Casimir operator should be rewritten as
\be
C(su(3))=H_1(H_1+2)+\frac{3}{4}H_2^2+E_{-1}\star E_1+E_{2}\star E_{-2}+E_{-3}\star E_3.
\ee
Then we can evaluate it to obtain
\be
C(su(3))\star f_{10}(x)=3f_{10}(x).
\ee
The same value can be obtained from other states of the octet by rearranging the order of $E_i$ and $E_{-i}$, so that it fits to the conditions at vertex of the hexagon.
%
%
\section{Unification of Quantizations in the External space and Internal Space}
In section 3, we quantize a Hamiltonian of one-dimensional harmonic oscillator in the two dimensional phase space in terms of the star product. We apply the star product method also to the Lie algebra. In section 4, we quantize the Lie algebra $su(2)$ in the three dimensional representation space, and Lie algebra $su(3)$ in the eight dimensional representation space. The quantization is made by setting CRs among the coordinates in each dimension. Once the CRs among the coordinates are set, or explicit $\Theta^{\mu\nu}(x)$ are determined, the quantization is implemented. The terminology ``quantization" is meant by not only the quantization of the energy levels but also computing quantum numbers in the Lie algebra. As far as quantization by using the star product is concerned, there seems to be no reason to distinguish both quantizations in the phase space and representation space. Only difference between them is the choice of entities of $\Theta^{\mu\nu}(x)$ in Eq.(\ref{qoe}). 

Then it is natural to unify two quantizations. The unification of the external space and the internal space reminds us of the Kaluza-Klein(K-K hereafter) theory\cite{KK}. In the K-K theory, the unification of the external space and internal space in higher dimension together with the dimensional reduction leads to the 4-dimensional theory with internal symmetries, like the Einstein gravity with a $U(1)$ gauge field. The K-K theory is a classical theory. In this paper, we claim that the quantizations are unified in higher dimensions by setting the CRs in higher dimensions, which brings about the energy levels in the external phase space and the quantum numbers in the representation space, at the same time. 
   
We show this unification idea by a toy model in which the external phase space is two dimensional phase space and the internal representation space is three dimensional. Then we discuss the  quantization in terms of the star product in five dimensions. We assume that the internal and external spaces are factorized. The star product is characterized by $\Theta^{\mu \nu}(x),~(\mu,\nu=1,2,\cdots,5)$ given by
\be
(\Theta^{\mu \nu}(x))=\theta
\pmatrix{
0	& r	&0	&0	&0\cr
-r  	&0	&0	&0      &0\cr
0	&0	&0 	& x^5 	& -x^4\cr
0	&0	&-x^5 	& 0	& x^3\cr
0	&0	&x^4	&-x^3	& 0
},
\ee
where $r$ is some constant. Although we can not determine the magnitude of $\theta$ and $r$ separately, we might regard their multiplication $\theta \cdot r$ as $\hbar$.  We factorize the Wigner function $f(x)$ in $2+3$ dimensions into two parts, one in the external phase space $f_1(x^1,x^2)$ and the other in the internal representation space. We assume that the Wigner function of five variables is factorized to the  Wigner function $f_1(x^1,x^2)$ of two variables and the Wigner function $f_2(x^3,x^4,x^t)$ of three variables
\be
f(x)=f_1(x^1,x^2)\cdot f_2(x^3,x^4,x^t).
\ee
Then the remaining computation has been already shown in previous sections. When we assume CRs in higher dimensions, the stargen function $f_1(x^1,x^2)$ gives rise to the states for the quantized energy spectrum,  and the stargen function $f_2(x^3,x^4,x^t)$ yields the quantum numbers of su(e) like  an isospin state. In this way, the energy spectrum and the isospin state, which is either upstate or downstate of $su(2)$. These quantum numbers emerge simultaneously by the quantization in higher dimensions. Although this is simply a toy model, we can extend the present model in a more realistic way.

%
%
\section{Summary and Discussion}

In this paper, we have shown the star product construction of CR given by (\ref{qoe}), which includes the $\theta$-deformation quantization and the $x$-dependent $\Theta^{\mu \nu}(x)$ cases like the Lie algebra. In the curved space time, where the coordinates are not perpendicular to each other and so the vielbein $e^{\mu}_i(x)$ is not $\delta^{\mu}_i$, it seems natural to use $x$-dependent $ \Theta^{\mu \nu}(x)$. We showed that the CR is reduced to that of $su(2)$ or $su(3)$ when $\Theta^{\mu \nu}(x)$ has special entities. A special case of constant $\Theta^{\mu \nu}(x)$ is nothing but the $\theta$-deformation. There is no essential difference between the quantization of quantum mechanics and the quantization of the Lie algebra from the viewpoint of the star product quantization. Then, it seems natural to unify the quantizations in the internal space and external phase space to higher dimensions. In section 5, we gave a toy model of unified quantization where one-dimensional harmonic oscillator and $su(2)$ internal symmetry are simultaneously quantized. 

In section 4 we discuss the quantization of $su(2)$ and $su(3)$. When we compare two quantizations, $su(2)$ is much easier because the rank is just one. In order to confine states in a multiplet of $su(2)$, we limited outward translations from the multiplet. This limitation corresponds to the vacuum condition in the quantum mechanics like the one-dimensional harmonic oscillator. In $su(2)$, one limitation brings about another limitation by complex conjugation. Therefore, the number of states is finite. On the other hand, in the $\theta$-deformation the number of states is infinite.

Though the model in section 5 is just a toy model, we can generalize it to more realistic model that is comprised of realistic external phase space and realistic internal symmetry space. Since a Wigner function expresses a probability of the corresponding state, it would be possible to compute the possibilities of various dimensional reductions when we obtain all the Wigner functions corresponding to those dimensional reductions. For example, it might be possible to compare the magnitude relation of probabilities of $su(3)$ and $su(2)\times u(1)$ by use of the Wigner functions. In the present discussion, we made use of the algebraic method in finding the energy spectrum and quantum numbers of the Lie algebra invariants. But we can also use the analytic method in finding those quantum numbers. Then we need to solve the stargen-value equation to obtain the stargen functions in this analytic method. These functions are nothing but the Wigner functions representing the probabilities. A study investigating in this direction will be reported in the future.

\end{document}